\documentclass[12pt]{article}
\let\a=\alpha\let\b=\beta\let\d=\delta

\newcommand{\nn}{\nonumber}

\newcommand{\be}{\begin{equation}}
\newcommand{\ee}{\end{equation}}
\newcommand{\bea}{\begin{eqnarray}}
\newcommand{\eea}{\end{eqnarray}}
\newcommand{\eps}{\epsilon}

\renewcommand{\paragraph}[1]{
\vspace{.8mm}\par\noindent {\sl #1}\\
\vspace{0.2mm} }

\newcommand{\eqn}[1]{(\ref{#1})}
\newcommand{\ft}[2]{{\textstyle\frac{#1}{#2}}}

\newcommand{\ba}{\left(\begin{array}}
\newcommand{\ea}{\end{array}\right)}

\def\da{{\dot\alpha}}
\def\db{{\dot\beta}}
\def\dc{{\dot\gamma}}

\newsavebox{\uuunit}
\sbox{\uuunit}
{\setlength{\unitlength}{0.825em}
 \begin{picture}(0.6,0.7)
\thinlines
\put(0,0){\line(1,0){0.5}}
\put(0.15,0){\line(0,1){0.7}}
 \put(0.35,0){\line(0,1){0.8}}
\multiput(0.3,0.8)(-0.04,-0.02){12}{\rule{0.5pt}{0.5pt}}
\end {picture}}
\newcommand {\unity}{\mathord{\!\usebox{\uuunit}}}
\begin{document}
\begin{titlepage}
\begin{flushright}
SU-ITP-99/27\\
KUL-TF-99/21\\
{\tt hep-th/9906118}\\
June 15, 1999\\
\end{flushright}
\vskip 2cm
\begin{center}
{\large {\bf A Simple Particle Action from a\\[5pt]
Twistor Parametrization of AdS$_5$}}
\vskip 1.5cm
{\bf  Piet Claus$^{\dagger a}$, J.~Rahmfeld$^{* b}$ and Yonatan Zunger$^{* c}$}\\
\vskip.5cm
{\small
$^\dagger$
Instituut voor theoretische fysica, \\
Katholieke Universiteit Leuven, B-3001 Leuven, Belgium\\
\par
\
\par
$^*$ Physics Department, \\
Stanford University, Stanford, CA 94305-4060, USA
}
\vskip 2cm
\begin{abstract}

The $SO(4,2)$ isometries of $AdS_5$ are realized non-linearly on its
horospherical coordinates $(x^m,\rho)$.  On the other hand, Penrose
twistors have long been known to linearly realize these symmetries on
4-dimensional Minkowski space, the boundary of $AdS_5$, parametrized by
$x^m$.  Here we extend the twistor construction and define a pair of
twistors, allowing us to include a radial coordinate in the construction.
The linear action of $SO(4,2)$ on the twistors induces the correct
isometries of $AdS_5$.  We apply this new construction to the study of the
dynamics of a massive particle in $AdS_5$.  We show that in terms of the
twistor variables the action takes a simple form of a
1-dimensional gauge theory.  Our result might open up the possibility to
find a simple worldvolume action also for the string propagating on
$AdS_5$.

\end{abstract}
\end{center}

\vfill
\footnoterule \noindent
{\footnotesize $\phantom{a}^a$ e-mail: piet.claus@fys.kuleuven.ac.be.}\\
{\footnotesize $\phantom{b}^b$ e-mail: rahmfeld@leland.stanford.edu.}\\
{\footnotesize $\phantom{c}^c$ e-mail: zunger@leland.stanford.edu.}
\end{titlepage}

In \cite{CGKRZ} a two-dimensional world-sheet model was considered
with manifest $SU(2,2|4)$ symmetry. The basic variables of the model
are supertwistors which realize the symmetry linearly. From the
supertwistors one can derive 4 $x^m$ and 16 $\theta$ \cite{Penrose,Ferber}
coordinates on which $SU(2,2|4)$ is realized as the non-linear
superconformal symmetries.
Since the same symmetry group also appears as the isometry group
of the $AdS_5\times S^5$ superspace, it is natural to wonder
in view of the $AdS$/CFT correspondence \cite{malda}, whether
supertwistors might be useful variables to describe strings
in this background.
In this brief note, we extend the purely bosonic twistor construction
to include a radial coordinate, based on the coset construction of $AdS_5$.
The linear action of the $SU(2,2)$ group on these twistors induces the correct
isometries of $AdS_5$. As a warm-up to applying  these new variables
to strings propagating in an $AdS_5\times S^5$ superspace, we then consider
the (non-linear) action of a massive particle on $AdS_5$ and show how
the new twistors bring the action into a simple (and quantizable
\cite{BRSTpart}) form,
generalizing the work of \cite{Townsend}. In fact, the resulting theory
is nothing but a one-dimensional $U(1)\times SU(2)$ gauge theory of the form
\begin{equation}
S = -i \int d\tau\, \left(\bar {\cal Z}^I D_{\tau} {\cal Z}^I + 2 i u^0 m
R\right)\,,
\end{equation}
where
\begin{equation}
D_\tau {\cal Z}^I = (\delta^{IJ}\partial_\tau - i u^a t^{IJ}_a) {\cal
Z}^J\, ,
\end{equation}
and the $t^{IJ}_a$ matrices are $U(1)\times SU(2)$ generators. The
rest of the notation will be explained in the main body.

\bigskip

We begin by recalling the $AdS_5$ isometries, their relation to
the standard twistor formalism in the boundary limit, and
the general construction of coset isometries.
$AdS_5$ space in horospherical coordinates $\tilde x^{\tilde m} =
(x^m,\rho)$ is described by the metric
\begin{equation}
ds^2 = d\tilde x^{\tilde m} \tilde g_{\tilde m\tilde n} d\tilde x^{\tilde n} =
e^{m} \eta_{mn} e^{n} + e^\rho e^\rho =
\rho^2 dx^2 + R^2 \,\left(\frac{d\rho}\rho\right)^2\,,
\label{adsmetric}
\end{equation}
where $R$ is the radius of curvature.  The isometries and compensating
Lorentz transformations $\ell$ (acting on $e^m, e^\rho$) \cite{CK} are
given by
\begin{eqnarray}
\delta x^m &=& - \left(a^m + \lambda_M^{mn} x_n + \lambda_D x^m +
x^2 \Lambda_K^m - 2 x\cdot \Lambda_K x^m\right) -
R^2\, \left(\frac{\Lambda_K^m}{\rho^2}\right)\,,   \nonumber\\
\delta \rho &=& (\lambda_D - 2 x \cdot \Lambda_K)\rho\,,\nonumber\\
\ell^{mn}(\tilde x) &=& \lambda^{mn}_M - 4 x^{[m} \Lambda_K^{n]}\,,
\nonumber\\
\ell^{m\rho}(\tilde x) &=& -2 R\, \frac{\Lambda_K^m}{\rho}\,,
\label{adsisom}
\end{eqnarray}
where $a^m$, $\lambda_M^{mn}$, $\lambda_D$, and $\Lambda_K^m$ are the
constant parameters of translations, Lorentz boosts, dilatation, and special
conformal transformations.

In the limit $\rho\rightarrow\infty$ the radial coordinate decouples
from the transformations and one is left with the conformal transformations
of 4-dimensional Minkowski space. These non-linear transformations are
nicely encoded in the
transformation properties of a twistor\footnote{As usual in the twistor
literature, $\lambda_\alpha$ and $\bar\mu^\da$ are commuting, two-component
complex spinors. Our conventions are explained in the appendix.}
\begin{equation}
{\cal Z}\equiv\ba{c}\lambda_\alpha\\\bar\mu^{\da}\ea\,,
\label{twist}
\end{equation}
transforming under {\sl rigid} transformations $g$ in the fundamental of
$SU(2,2)$\footnote{The universal covering group of the
conformal group $SO(4,2)$ is $SU(2,2)$.  Its compact $SU(2)$ subgroups act as
rotations on chiral and antichiral spinors, respectively. }
\be
\delta  \ba{c}\lambda_\alpha\\\bar\mu^{\da}\ea =
- g\ba{c}\lambda_\alpha\\\bar\mu^{\da}\ea \equiv
\pmatrix{L_\alpha{}^\beta+\frac{1}{2}
D \delta_\alpha{}^\beta & iK_{\alpha\dot\beta} \cr
i A^{\dot\alpha\beta} & -\bar
L^{\dot\alpha}{}_{\dot\beta} - \frac{1}{2}
D\delta^{\dot\alpha}{}_{\dot\beta}}
\ba{c}\lambda_\beta\\\bar\mu^{\db}\ea \, ,
\label{VectorTwistor}
\ee
where we used the spinorial notation (see the appendix)
\begin{eqnarray}
x^{\da\a} &=& \frac{x^m}R (\bar\sigma_m)^{\da\a}\,,\nonumber\\
A^{\da\a} &=& \frac {a^m}R (\bar \sigma_m)^{\da\a}\,,\qquad
K_{\a\da}\ =\ R\, \Lambda_K^m (\sigma_m)_{\a\da}\,, \qquad
D \ =\ \lambda_D\,,\nonumber\\
L_\a{}^\b &=& \frac14 \lambda_M^{mn} (\sigma_{mn})_\alpha{}^\b\,,
\qquad \bar L^{\da}{}_\db \ =\ -\frac 14 \lambda_M^{mn}
(\bar \sigma_{mn})^\da{}_\db\,,
\end{eqnarray}
i.e.  we choose to take all our variables dimensionless.
The {\sl conformal twistor relation} between the two spinors $\lambda$
and  $\bar \mu$,
\be
\bar\mu^{\da}=-i x^{\da \a}\lambda_\a \, ,
\label{RelationConf}
\ee
then induces the non-linear conformal transformations of $x$ from the simple
linear transformation of the twistor ${\cal Z}$.  The non-trivial step in
this parametrization of conformal space is the twistor relation
\eqn{RelationConf}, which can be derived from the dynamics
of a massless particle \cite{Ferber}.  However, here we derive
the twistor relation from a purely algebraic perspective, as the
twistor parametrization of conformal space is closely related to the
coset-space construction. This will allow us to generalize the procedure
to the full $AdS_5$ space.

The parametrization of a coset space is encoded in the
coset representative $v(X)$, where $X$ are the coordinates of the space.
As a function of coordinates $v(X)$ transforms under an
infinitesimal action of the group $G$ as
\be
\delta v(X)\equiv - \xi^m_g (X) \partial_m v(X)= - g v(X) + v(X) h_g(X)
\label{isomv}
\ee
with $m=1,..., \dim G/H$, $g\in {\bf G}$, the Lie-algebra associated to $G$,
and $h_g(X)$ is a space-time dependent compensating transformation in the
Lie-algebra of $H$. Eq.~\eqn{isomv} is equivalent to the general relation
\cite{CK}
\begin{equation}
v^{-1} g v = \xi_g^m L_m + h_g(X)\,,
\end{equation}
where $L_m = v^{-1} \partial_m v$ is the Cartan 1-form in {\bf G}.
The $\xi^m_g$ are the isometries, $\delta X^m = -
\xi_g^m(X)$, that realize $G$ non-linearly on the coordinates.
Clearly, an object
\begin{equation}
{\cal Z}=v(X){\cal Z}_0 \,
\label{vector}
\end{equation}
will transform rigidly in the fundamental of $G$, i.e.
$\d {\cal Z}=-g{\cal Z}$, iff $\d {\cal Z}_0=- h_g(X) {\cal Z}_0$.

Twistors turn this reasoning upside down. Whenever we define a vector
${\cal Z}_0$ by \eqn{vector} and  demand that ${\cal Z}$ transforms in the
fundamental of $G$, it is guaranteed that ${\cal Z}_0$ will transform under
the correct compensating transformation $h_g(X)$. If we want to impose a
twistor relation on ${\cal Z}$, it turns out to be most useful to put a
constraint on ${\cal Z}_0$, effectively relating the components of ${\cal Z}$
via $X$. Hereby, one only has to demand that the constraint is covariant under the
compensating $H$-transformations. A consistent twistor relation can therefore be
obtained from the coset space by defining a vector ${\cal Z}_0$ through
\eqn{vector} and constraining it in a $H$ covariant way.

We illustrate the above discussion with the twistor formulation of conformal
space. The conformal transformations in spinorial notation
\begin{equation}
\delta x^{\da\a} = -\left( A^{\da\a} + x^{\da\b} L_\b{}^\a + \bar
L^{\da}{}_{\db} x^{\db \a} + D x^{\da\a} + x^{\da\b} K_{\b\db}
x^{\db\a}\right)\, ,
 \label{conformal}
\end{equation}
induces the compensating $H$ transformation
\begin{equation}
h_g^{Conf} = \ba{cc}
-L_\alpha{}^\beta - \frac D2 \delta_\a{}^\b - K_{\a\da} x^{\da\b} & -i
K_{\a\da} \\
0 &\bar L^{\da}{}_\db + \frac D2 \delta^\da{}_\db - x^{\da\b} K_{\b\db}
\ea\,.
\label{confcomp}
\end{equation}
These transformations are simply the isometries of the coset
space $SU(2,2)/ISO(1,3) \times D$, with the coset representative
\begin{equation}
v^{Conf}(x) =  \ba{cc} \d_\a{}^\b   & 0\\
                    -i x^{\da\a} & \d^\da{}_\db\ea\,.
\end{equation}
{}From \eqn{twist} and \eqn{vector} one constructs
\begin{equation}
{\cal Z}_0 \equiv \ba{c} \lambda_0{}_\alpha \\ \bar \mu_0{}^\da\ea
= \ba{c} \lambda_\a \\ i x^{\da\a} \lambda_\a + \bar \mu^\da\ea\, ,
\label{Z0confuc}
\end{equation}
which needs to be constrained $H$-covariantly. The only
form of ${\cal Z}_0$ transforming covariantly under \eqn{confcomp} is
\be
{\cal Z}_0=\pmatrix{\lambda_0{}_\a\cr  0},
\label{conformalZ0}
\ee
which implies the conformal twistor relation \eqn{RelationConf}, as desired.
In this case finding the proper constraint (\ref{conformalZ0})
was straightforward. The $SO(1,3)$ in $H$ allows for chiral/anti-chiral
spinors, or Majorana spinors. However, only the chiral spinor
$\lambda_0^\alpha$ of (\ref{conformalZ0}) transforms covariantly under
special conformal transformations.

This example illustrates nicely the method we have developed of constructing
twistors via the coset representative. We can now extend the procedure to the
full $AdS_5$ case.

The metric and isometries (\ref{adsmetric}) and (\ref{adsisom}) follow from
the coset construction of $AdS_5=SO(4,2)/SO(4,1)$. We choose
the coset representative $v^{AdS}(x^m,\rho)$ in the spinor representation of
$SO(4,2)$
\begin{equation}
v^{AdS}(x,\rho) = \ba{cc} \rho^{1/2} \d_\a{}^\b   & 0\\
                  -i \rho^{1/2} x^{\da\a} & \rho^{-1/2}\d^\da{}_\db\ea\,,
\end{equation}
and, in spinorial notation,  the isometries (\ref{adsisom}) take the form
\begin{eqnarray}
\delta x^{\da\a} &=& -\left( A^{\da\a} + x^{\da\b} L_\b{}^\a +
\bar L^{\da}{}_{\db} x^{\db \a} + D x^{\da\a} + x^{\da\b} K_{\b\db}
x^{\db\a}\right) -  K^{\da\a} \rho^{-2}\,,\nn\\
\delta \rho &=& \left(D + K^{\da\a} x_{\a\da}\right) \rho\,.
\label{eq:isom1}
\end{eqnarray}
The compensating $SO(1,4)$ transformation reads
\begin{equation}
h_g^{AdS} = \ba{cc}
- L_\a{}^\b - K_{\a\da} x^{\da\b} + \ft12 K_{\gamma\dc}
x^{\dc\gamma}\delta_\a{}^\b &
-i \rho^{-1} K_{\a\db}\\
i \rho^{-1} K^{\da\b} & \bar L^{\da}{}_\db - x^{\da\b} K_{\b\db} + \ft12
x^{\dc\gamma} K_{\gamma\dc} \delta^{\da}{}_\db \ea\,.
\label{eq:isom2}
\end{equation}

We wish to replace the coordinates $(x^m,\rho)$ with twistor variables, on
which a linear $SU(2,2)$ transformation induces (\ref{eq:isom1}) and
(\ref{eq:isom2}).  Naturally, one would want to define as above a
twistor ${\cal Z}$, and relate it via the coset representative
to a vector ${\cal Z}_0$. On ${\cal Z}_0$  then a constraint has
to be imposed transforming covariantly under $H$, effectively
relating the components of ${\cal Z}$ via the coset coordinates.
In this case we have $H=SO(1,4)$, and ${\cal Z}_0$ should transform
in the spinor representation of this group. There are
neither Weyl nor Majorana fermions in (1,4) dimensions, and there does
not exist a covariantly constrained spinor. Therefore,
a straightforward extension of (\ref{RelationConf}) incorporating
the radial $\rho$ coordinate of the $AdS$ space does not exist.
However, in (1,4) dimensions one can define pseudo-symplectic Majorana
fermions $\psi^I$ satisfying
\be
(\psi^I)^*=\epsilon ^{IJ} B\psi^J\,,\qquad I,J = 1,2\,.
\ee
The matrix $B$ satisfies
\be
B^T=-B, \qquad B^\dagger B=1,\qquad BB^* = -1\,,\qquad  \gamma^*_{\tilde m} =
-B \gamma_{\tilde m} B^{-1},
\ee
where $\gamma_{\tilde m}$ are the 5-dimensional $\gamma$-matrices.
In the two-component notation used above the matrix $B$ takes the form
\be
B=\pmatrix{0 & -\varepsilon_{\da\db}\cr \varepsilon^{\a\b} & 0}\,.
\ee
Since one can impose a restriction only on a
pair of spinors,  we have to extend the twistor construction to
a pair of twistors
\be
{\cal Z}^I=\pmatrix{\lambda_\a^I \cr \bar \mu^{\da I}}
\ee
transforming in the fundamental of $SU(2,2)$
\be
\d {\cal Z}^I= - g {\cal Z}^I\, .
\label{Fund}
\ee
Just as in (\ref{vector}) one relates these twistors to a
pair of spinors ${\cal Z}^I_0$
via
\be
{\cal Z}^I=v^{AdS}(x,\rho) {\cal Z}^I_0\, ,
\ee
where ${\cal Z}^I_0$ transforms as
\be
\d {\cal Z}_0^I=-h_g(X) {\cal Z}_0^I\, .
\ee
As discussed, the only $H$-covariant condition which
can be placed on the ${\cal Z}_0^I$ is
the pseudo-symplectic Majorana condition
\be
({\cal Z}_0^I)^*=\eps^{IJ}B{\cal Z}_0^J \, ,
\label{AdSCond}
\ee
which leads to the form
\be
{\cal Z}_0^I=\pmatrix{\lambda^I_{0\a} \cr
\eps^{IJ}\bar \lambda^{\da J}_0}
\ee
with unrestricted $\lambda^I_0$ and $\lambda_0^{\da I} =
(\lambda_0{}_\b^I)^* \varepsilon^{\db\da}$.
The generalization of (\ref{RelationConf}) now becomes
\be
\bar \mu^{\da I}=-i x^{\da \a}\lambda_\a^I+\frac{\eps^{IJ}}{\rho}
\bar \lambda^{\da J}\,.
\label{RelationAdS}
\ee
It is straightforward to verify that from this $AdS$ {\sl twistor relation},
(\ref{Fund}) and (\ref{VectorTwistor}), the $AdS$ isometries (\ref{eq:isom1})
follow.

\medskip

The pair of twistors trade the 5 coordinates $(x^m, \rho)$ for
8 complex degrees of freedom ${\cal Z}^I$, which are related as in
\eqn{RelationAdS}. This relation was derived from a purely algebraic
perspective. Using this explicit relation we can find the invariant
quadratic constraints the twistor pair satisfies. The $SU(2,2)$ group can
be defined as the transformations that leave the metric
\begin{equation}
H_{SU(2,2)} = \left(\begin{array}{cc}0 &\unity\\\unity& 0
\end{array}\right)
\end{equation}
invariant. Therefore the bilinear form
\begin{equation}
\bar {\cal Z}(1) {\cal Z}(2) = \mu^\alpha(1) \lambda_\alpha(2) +
\bar\lambda_\da(1)\bar\mu^\da (2)\,,
\end{equation}
where
\begin{equation}
\bar {\cal Z}={\cal Z}^\dagger H = (\mu^\alpha , \bar\lambda_\da) \,
\end{equation}
is the conjugate representation, is a quadratic invariant.
With a pair of twistors we can define the general bilinear invariant
\begin{equation}
{\cal Z}^I g^{IJ} {\cal Z}^J = {\cal Z}^I_0 g^{IJ} {\cal Z}^J_0\,,
\label{AdSinvariant}
\end{equation}
where $g^{IJ}$ is 2-dimensional constant complex matrix.
The equality above holds because ${\cal Z}^I$ and ${\cal Z}_0^I$ are
related by the $SU(2,2)$ matrix $v^{AdS}$. We can expand the matrix $g$
in a basis
\begin{equation}
t_a^{IJ} = \left\{\delta^{IJ}, (\hat \sigma_i)^{IJ}\right\}\,,
\label{U2matrices}
\end{equation}
where $\hat\sigma_i$ are the Pauli-matrices satisfying $\hat \sigma_1 \hat
\sigma_2 = i \hat \sigma_3$ and $a$ takes values $0,1,2,3$.
{}From the relation \eqn{RelationAdS} we obtain the constraints
\begin{equation}
\bar {\cal Z}^I \hat \sigma_i^{IJ} {\cal Z}^J = 0\,,
\label{twistorconstraints}
\end{equation}
which imply that the two components ${\cal Z}^1$ and ${\cal Z}^2$ of the
twistor pair have the same `length' $\bar {\cal Z}^1 {\cal Z}^1 = \bar{\cal
Z}^2 {\cal Z}^2$ and are `orthogonal' $\bar {\cal Z}^1 {\cal Z}^2 =
\bar {\cal Z}^2 {\cal Z}^1 = 0$.
However, there is a fourth independent invariant bilinear
\begin{equation}
\bar {\cal Z}^I \delta^{IJ} {\cal Z}^J = 2\, m R \,,
\label{scale}
\end{equation}
which takes arbitrary real values $m$. The parameter $m$ has the dimension
of a mass. Indeed, we will relate it to the mass of a particle propagating
in the $AdS_5$ background. Before discussing this we contrast these
constraints to the conformal space. In that case there is only one twistor
and the quadratic invariant
\begin{equation}
{\cal Z} {\cal Z} = {\cal Z}_0 {\cal Z}_0
\label{confsurface}
\end{equation}
necessarily vanishes due to \eqn{conformalZ0}. This equation defines a
surface in twistor space, which is related to the phase space of a massless
particle \cite{Ferber}.

The first order (or phase space) action of a `massive' particle in
$AdS_5$ is given by
\begin{equation}
S = \int d\tau \left[\tilde P_{\tilde m} \partial_\tau \tilde x^{\tilde m}
-\frac {e}{2} \left(\tilde P_{\tilde m} \tilde g^{\tilde m\tilde n} \tilde
P_{\tilde n} + m^2\right)\right]\, .
\label{particleaction}
\end{equation}
The momenta $\tilde P_{\tilde m} = (P_m, P_\rho)$ are
conjugate to $\tilde x^{\tilde m} = (x^m,\rho)$ and
$\tilde g^{\tilde m\tilde n}$ is the inverse $AdS_5$-metric \eqn{adsmetric}.
Translating this into two-component notation yields
\begin{equation}
S = -\int d\tau\,\left[ \frac12 P_{\a\da} \dot x^{\da\a} - P_\rho \dot \rho
- \frac{e}{2R^2}
\left( \frac{1}{2\rho^2} P_{\a\da} P^{\da\a} - \rho^2 P_\rho^2 -
m^2 R^2\right)\right]\,,
\end{equation}
where
\begin{equation}
P_{\a\da} = R P_m (\sigma^m)_{\a\da}\,.
\end{equation}
The worldline einbein $e(\tau)$ is a Lagrange multiplier for the mass-shell
constraint
\begin{equation}
\frac 1{2\rho^2} P_{\a\da} P^{\da\a} - \rho^2 P_\rho^2 =
m^2 R^2\label{massshell}\,.
\end{equation}
We rewrite the momenta in terms of the twistor variables $\lambda_\a^I$
introduced above by setting
\begin{eqnarray}
P_{\a\da} &=& 2 \lambda_\a^I \bar\lambda_\da^I\,,\nonumber\\
P_\rho &=& -\frac i{2\rho^2} \left(\varepsilon^{\a\b} \epsilon^{IJ}
\lambda_\a^I \lambda_\b^J - c.c.\right)\,,
\end{eqnarray}
and the second pair of twistor components $\mu^I$ are introduced
as in \eqn{RelationAdS}.  It follows straightforwardly that
\begin{equation}
\frac 1{2\rho^2} P_{\a\da} P^{\da\a} - \rho^2 P_\rho^2 =
\left(\frac 1{2\rho}\left(\varepsilon^{\a\b} \epsilon^{IJ}
\lambda_\a^I \lambda_\b^J + c.c.\right)\right)^2 =
\frac1{4} \left(\bar {\cal Z}^I \delta^{IJ} {\cal Z}^J\right)^2\,.
\end{equation}
The constraint \eqn{massshell} forces this to equal $m^2$ and therefore we
relate the invariant $\bar {\cal Z}^I \delta^{IJ} {\cal Z}^J$ to the mass
$m$ as in \eqn{scale}. In other words \eqn{scale} is equivalent to
the mass-shell constraint. It turns out that, like for the massless
particle in 1+3 dimensions \cite{Townsend}, the action for an on-shell
particle \eqn{massshell} can be rewritten entirely in twistor variables,
\begin{equation}
S = - \int d\tau\, \left(\frac12 P_{\da\a} \partial_\tau x^{\da\a} - P_\rho
\partial_\tau \rho\right)
= - i \int d\tau\, \bar {\cal Z}^I \partial_\tau{\cal Z}^I\,,
\label{twistoraction}
\end{equation}
where the twistor pair is subject to the four real constraints
\eqn{twistorconstraints} and \eqn{scale}
\begin{equation}
\phi_a = \bar {\cal Z}^I t_a^{IJ} {\cal Z}^J - 2 \delta_a{}^0 mR\,.
\label{primconstraints}
\end{equation}
If we wish to elevate the twistors to independent variables,
the constraints (\ref{primconstraints}) have to be imposed via Lagrange
multipliers $u^a$ and the total action becomes
\begin{equation}
S = - i \int d\tau\, \left(\bar {\cal Z}^I \partial_\tau {\cal Z}^I
-i u^a \phi_a\right) \,.
\label{twistoractionC}
\end{equation}
The appearence of the Lagrange multiplier terms generates gauge symmetries
which act on the fields as
\begin{eqnarray}
\delta {\cal Z}^I &=& i \xi^a(\tau)  t_a^{IJ} {\cal Z}^J\,,\nonumber\\
\delta u^0 &=& \partial_\tau \xi^0(\tau) \,,\nonumber\\
\delta u^i &=& \partial_\tau \xi^i(\tau)
               + 2i \varepsilon^{ijk} \xi^j(\tau) u^k\,,
\end{eqnarray}
where $\xi^a(\tau)$ are 4 local parameters and form an $U(1)\times SU(2)$
algebra. Therefore, we can rewrite the action \eqn{twistoractionC} as a
1-dimensional gauge theory action
\begin{equation}
S = -i \int d\tau\, \left(\bar {\cal Z}^I D_{\tau} {\cal Z}^I + 2 i u^0 m
R\right)\,,
\label{1dgauge}
\end{equation}
where
\begin{equation}
D_\tau {\cal Z}^I = (\delta^{IJ}\partial_\tau - i u^a t^{IJ}_a) {\cal
Z}^J\,
\end{equation}
and the Lagrange multipliers act as gauge fields.

\bigskip

We have shown that the construction of twistors, realizing the
conformal symmetry $SU(2,2)$ of 4-dimensional Minkowski space linearly,
can be generalized to include the radial coordinate of the $AdS_5$ space.
To do so, we employed the coset descriptions of the conformal
space and the $AdS_5$ space, and we had to extend the twistor
construction to a pair of twistors subject to important
constraints. In essence, we built a bridge between the twistor formulation
of $SU(2,2)$, and the description of $AdS_5$ as a hyperboloid in (2,4)
dimensions. In the latter, the coordinates
of the 6-dimensional embedding space also realize the conformal
symmetry $G=SO(2,4)\sim SU(2,2)$ linearly, although in the
fundamental of $SO(2,4)$. The generalized twistors realize the
symmetry linearly in the spinor representation of $SO(2,4)$ which
is equivalent to the fundamental representation of $SU(2,2)$.
One advantage of our procedure is that it
seems to have a straightforward generalization to the supersymmetric case
$SU(2,2|{\cal N})$, i.e. to find a twistor realization of the $AdS_5\times
S^5$ superspace. All one needs to do is to construct the
supercoset representative and find the proper constraint on
${\cal Z}_0$. This will essentially be the same constraints as for the bosonic space
since the compensating transformation in the supersymmetric case is also
purely bosonic. In that way one could find the supersymmetric
generalization of defining $AdS_5 \times S^5$
as a 10-dimensional hypersurface in 12 dimensions.

We have also shown that these twistors appear naturally as the fundamental
variables of a massive particle propagating in $AdS_5$.  It would be
interesting to investigate whether we can extent the superparticle action
to a simple supertwistor one as well.

The most exciting aspect of our result however is the simplicity of the
action \eqn{1dgauge}.  The quantization of this action will be considered
in an accompanying paper \cite{BRSTpart}.  It might pave the way to a
simple quantization of the string on $AdS_5$.  One would hope that
(\ref{twistoraction}) generalizes to the string action in the same
straightforward way it did in flat space, viewed as the boundary limit of
$AdS_5$ (\cite{Townsend} vs.~\cite{CGKRZ}).  If this is the case then the
string spectrum constructed in \cite{CGKRZ} from a stringy toy world-sheet
model should indeed be the spectrum of strings on $AdS_5\times S^5$.

\bigskip
\bigskip

\noindent
{\bf Acknowledgements:} We enjoyed very useful discussions with Renata
Kallosh.  The work of P.C.~was supported by the European Commission TMR
program ERBFMRX-CT96-0045.  J.R.~and Y.Z.~were supported in part by NSF
grant PHY-9870115.
\newpage
\noindent
{\bf \large Spinor notation}

\medskip

\noindent
Here we collect some useful formulae and fix notation.  We define the
$\sigma$-matrices as
\begin{equation}
\sigma_{m} = \{\unity,\hat \sigma_i\}\,,\qquad \bar \sigma_{m} = \{\unity,
- \hat\sigma_i\}\,, \qquad \sigma_{mn} = \sigma_{[m} \bar\sigma_{n]}\,,
\qquad \bar\sigma_{mn} = \bar\sigma_{[m} \sigma_{n]}\,,
\end{equation}
where $\hat\sigma_i$ are the conventional Pauli-matrices satisfying $\hat
\sigma_1 \hat \sigma_2 = i \hat\sigma_3$.  They have the index structure
\begin{equation}
(\sigma_m)_{\a\da}\,,\quad (\bar \sigma_m)^{\da\a}\,,\quad
(\sigma_{mn})_\a{}^\b\,,\quad (\bar \sigma_{mn})^\da{}_\db\,.
\end{equation}
Undotted indices are always contracted NW-SE, while dotted indices are
contracted SW-NE.  Indices are contracted with an $\varepsilon$-tensor,
defined by
\begin{equation}
\varepsilon_{12} = - \varepsilon_{\dot1\dot2} = \varepsilon^{12} =
-\varepsilon^{\dot1\dot2} = 1\,,\qquad \varepsilon^{\a\gamma}
\varepsilon_{\b\gamma} = \delta^\a{}_\b\,, \qquad \varepsilon_{\da\dc}
\varepsilon^{\db\dc} = \delta_{\da}{}^\db\,.
\end{equation}
This implies the relationships
\begin{equation}
(\sigma_m)_{\a\da} = (\sigma^T_m)_{\da\a} = \varepsilon_{\da\db}
(\bar\sigma_m)^{\db\b} \varepsilon_{\b\a}\,,\qquad (\bar \sigma_m)^{\da\a}
= (\bar \sigma^T_m)^{\a\da} = \varepsilon^{\a\b} (\sigma_m)_{\b\db}
\varepsilon^{\db\da}\,,
\end{equation}
We also have the identities
\begin{equation}
\sigma_{(m} \bar\sigma_{n)} = \bar \sigma_{(m} \sigma_{n)} = -\eta_{mn}
\unity\,,
\end{equation}
where $\eta = \mbox{diag}(-,+,+,+)$.  To translate between spinor notation
and vector notation we associate a real 4-dimensional vector $X^m$ to a
hermitian matrix
\begin{equation}
X^{\da\a} = X^m (\bar \sigma_m)^{\da\a}\,\quad \mbox{and} \quad X_{\a\da}
= X^m (\sigma_m)_{\a\da}\,\qquad\rightarrow\qquad X_{\a\da} =
\varepsilon_{\da\db} X^{\db\b} \varepsilon_{\b\a}\,.
\end{equation}
In the same way we translate a real anti-symmetric tensor $T^{mn}$ to the
pair of chiral-antichiral matrices
\begin{equation}
T_\alpha{}^\beta = \frac14 T^{mn} (\sigma_{mn})_\a{}^\b \,\quad \mbox{and}
\quad \bar T^\da{}_{\db} = -\frac14 T^{mn} (\bar\sigma_{mn})^\da{}_\db\,.
\end{equation}
which are related by hermitian conjugation.
\par
The 5-dimensional $\gamma$-matrices are constructed out of $\sigma, \bar
\sigma$ by
\begin{equation}
\gamma_m = i \ba{cc} 0 & \sigma_m \\ \bar\sigma_m & 0\ea\,, \qquad
\gamma_\rho \equiv \gamma_5 = i \gamma_0 \gamma_1 \gamma_2 \gamma_3 =
\ba{cc} -\unity & 0 \\ 0 & \unity \ea\,.
\end{equation}

\end{document}